\documentclass[onecolumn,11pt]{IEEEtran}
\usepackage{amsmath,amssymb}
\usepackage{graphicx}
\usepackage{epstopdf}
\usepackage{pgf}
\usepackage{tikz}
\usepackage{cite}
\usepackage{setspace}
\usepackage{hyperref}
\usepackage{pgffor}
\usepackage{pgfplots}
\usepackage{pgfplotstable}
\usetikzlibrary{arrows,shapes,positioning,plotmarks,fit}
\usepgflibrary{shapes.geometric}
\newcommand{\norm}[1]{\left\|{#1}\right\|}

\newtheorem{theorem}{Theorem}

\newtheorem{lemma}{Lemma}
\newtheorem{definition}{Definition}

\makeatletter
\def\mod#1{\allowbreak\mkern6mu{\operator@font mod}\,\,#1}
\makeatother

\title{Discrimination on the Grassmann Manifold: Fundamental Limits of Subspace Classifiers}

\author{
\IEEEauthorblockN{Matthew~Nokleby, {\em Member, IEEE}, Miguel Rodrigues, {\em Member, IEEE}, and Robert Calderbank, {\em Fellow, IEEE}} \thanks{
This work supported in part by AFOSR grant FA9550-13-1-0076 and the Royal Society International Exchanges Scheme IE120996. Preliminary elements of this work were presented at the IEEE Information Theory Workshop, Sept. 2013 and the IEEE Symposium on Information Theory, July 2014. Matthew Nokleby and Robert Calderbank are with Duke University, Durham, NC (emails: \{matthew.nokleby,robert.calderbank\}@duke.edu), and Miguel Rodrigues is with University College London (email: m.rodrigues@ucl.ac.uk)}}

\begin{document}
\maketitle

\begin{abstract}
We present fundamental limits on the reliable classification of linear and affine subspaces from noisy, linear features. Drawing an analogy between discrimination among subspaces and communication over vector wireless channels, we propose two Shannon-inspired measures to characterize asymptotic classifier performance. First, we define the {\em classification capacity}, which characterizes necessary and sufficient conditions for the misclassification probability to vanish as the signal dimension, the number of features, and the number of subspaces to be discerned all approach infinity. Second, we define the {\em diversity-discrimination tradeoff} which, by analogy with the diversity-multiplexing tradeoff of fading vector channels, characterizes relationships between the number of discernible subspaces and the misclassification probability as the noise power approaches zero. We derive upper and lower bounds on these measures which are tight in many regimes. Numerical results, including a face recognition application, validate the results in practice.
\end{abstract}


\section{Introduction}
The classification of high-dimensional signals arises in a host of situations, from face and digit recognition \cite{adini:PAMI97,hull:PAMI94,lecun:ICANN95} to tumor classification \cite{ross:Nature00,alon:PNAS99}, and to the music-identification app Shazam \cite{shazam}. These problems involve massive data sets---images with millions of pixels, DNA arrays with thousands of genes, or audio clips with tens of thousands of samples---which presents a substantial burden of computation and storage. Frequently, however, the data lie near a low-dimensional subspace of ambient space. For example, images of an individual's face, subject to constraints on pose, lighting, and convexity, lie almost entirely on a subspace of five to nine dimensions, regardless of the ambient dimension of the image \cite{georghiades:PAMI01,lee:PAMI05,epstein:PBMCF:95}. One therefore can pose classification tasks like face recognition as subspace classification problems.

When identifying low-dimensional subspaces, one can reduce the computation and storage burden by classifying from a low-dimensional representation of the signal of interest. This process is called {\em feature extraction}, and standard techniques, including linear discriminant analysis (LDA) and principal component analysis (PCA) \cite{fisher:AOE36}, as well as their myriad variations, are well studied. One pays a price, however, for computational tractability. In principle, extracting low-dimensional features from high-dimensional signals degrades classifier performance, and it is unclear {\em a priori} how many features are necessary to ensure success.

In this paper, we present a rigorous, information-theoretic characterization of classifier performance of high-dimensional data from low-dimensional features. We show that performance depends on several factors, including the number of subspaces to be discriminated, the number of features extracted, and the underlying subspace structure. In particular, we consider the classification of $k$-dimensional linear and affine subspaces of $\mathbb{R}^N$ from $M$ linear features corrupted by Gaussian noise.
To characterize classifier performance, we define two performance measures:
\begin{itemize}
	\item The {\em classification capacity}, which characterizes the number of unique subspaces that can be discerned as a function of the noise power, $N$, $M$, and $k$, as the latter three quantities approach infinity. Just as the usual Shannon capacity captures the phase transition of the error probability, as a function of the information rate, as the code length goes to infinity, the classification capacity captures the phase transition of the misclassification probability, in terms of the (logarithm of the) number of subspaces, as the signal dimension goes to infinity.
	\item The {\em diversity-discrimination tradeoff} (DDT), which characterizes the relationship between the number of subspaces and the misclassification probability as the noise power goes to zero. Just as the diversity-multiplexing tradeoff for fading vector wireless channels \cite{zheng:IT03} characterizes the number of codewords and the error probability in terms of a region of achievable exponent pairs in the signal-to-noise ratio (SNR), the DDT specifies a region of achievable exponent pairs in the noise power for the number of subspaces and the misclassification probability.
\end{itemize}

The motivation for the preceding definitions is an analogy between classification from noisy features and communication over non-coherent vector channels. Indeed, the title of our paper alludes to \cite{zheng:IT02}, which investigates the capacity of the block-fading non-coherent channel in geometric terms. It shows that, at high SNR and for sufficiently long coherence time, transmitters achieve near-capacity rates by sending {\em subspaces} as codewords. Therefore the decoding task is to discern subspaces from noisy observations, and the capacity corresponds to asymptotic packings in the Grassmann manifold. Further works give tighter bounds on the capacity and explore the diversity-multiplexing tradeoff of the non-coherent channel \cite{zheng:IT02,marzetta:IT99,zheng:Allerton02,yang:JSAC13}.

For the classification of $k$-dimensional linear subspaces from noisy, linear features, one can demonstrate a syntactic duality with communications over non-coherent vector channels. Specifically, the classification problem is dual to a non-coherent communications over a channel with $k$ transmit antennas, a single receive antenna, and a coherence time of $M$. In a preliminary version of this work, we applied results from \cite{zheng:IT02,zheng:Allerton02} directly to prove necessary conditions for successful classification \cite{nokleby:ITW13}.Ê These bounds translate into upper bounds on the classification capacity and diversity-discrimination tradeoff considered in this paper.

However, these bounds are somewhat crude. In the dual communications problem, the optimal transmission strategy employs only a single transmit antenna, which is equivalent to classifying subspaces of dimension $k=1$. Therefore, the upper bounds are loose when classifying subspaces of higher dimension. Furthermore, because the classification problem is not known to be {\em information stable} \cite{dobrushin:AMS63}, the mutual information between subspaces and features does not lower bound the classification capacity even for $k=1$. To prove tighter upper bounds on performance, we develop new bounds on the mutual information, and to prove lower bounds on performance we analyze the misclassification probability directly.

\subsection{Summary of Results}
Our primary contributions are upper and lower bounds, which are tight in many regimes, on the classification capacity and the diversity-discrimination tradeoff.

In Section \ref{sect:capacity}, we study the classification capacity. First we consider the classification of linear subspaces, which we model by taking the classes to follow zero-mean Gaussian distributions with approximately low-rank covariances. The covariances have two components: a rank $k$ component corresponding to the class subspace, and an identity component scaled by $\sigma^2$ corresponding to deviations from the subspace. We further suppose a prior distribution the subspaces which is uniform over the Grassmann manifold of $k$-dimensional subspaces in $\mathbb{R}^N$. We present an upper bound on the classification capacity, showing almost surely that the number of subspaces cannot scale any faster than $(1/\sigma^2)^{\frac{M-k}{2}}$. This result is intuitive: The lower the inherent signal dimension, the fewer features are required to classify the signal reliably. We also present a lower bound on the classification capacity, showing that the misclassification probability decays to zero, except for a set of subspaces having vanishing probability, provided the number of subspaces grows slower than $(1/\sigma^2)^{\frac{\min\{k,M-k\}}{2}}$. When $M \leq 2k$, the bounds are tight up to a $O(1)$ term. Furthermore, based on simulations presented in Section \ref{sect:experiments}, we conjecture that the upper bound is tight and that the gap between lower and upper bounds when $M > 2k$ is merely an artifact of the analysis.

We then consider the classification of {\em affine} subspaces, or linear subspaces translated by nonzero points. We model affine spaces by taking the classes to again be modeled by approximately rank-$k$ covariances, but this time to have nonzero means. We again suppose a uniform prior over the Grassmann manifold, and we further suppose that the means are distributed according to a standard Gaussian distribution. We characterize the classification capacity up to a $O(1)$ term, showing that the number of subspaces growing no faster than $(1/\sigma^2)^{\frac{M-k}{2}}$ is both necessary and sufficient for the probability of error to decay to zero, again except for a set of subspaces having vanishing probability.

In Section \ref{sect:ddt}, we study the diversity-discrimination tradeoff. For linear subspaces, we derive an upper bound, showing that the average misclassification probability decays no faster than $(1/\sigma^2)^{-\frac{\min\{k,M-k\}}{2}}$ as $\sigma^2 \to 0$ and that the misclassification probability exhibits an error floor when the number of subspaces scales faster than $(1/\sigma^2)^{\frac{M-k}{2}}$. We also derive a lower bound, showing that the misclassification capacity decays at least as $(1/\sigma^2)^{-\frac{\min\{k,M-k\} - r}{2}}$ when the number of subspaces scales as $(1/\sigma^2)^{\frac{r}{2}}$.  For affine spaces, we specify the DDT exactly, showing that the average probability decays as $(1/\sigma^2)^{-\frac{M-k-r}{2}}$ when the number of subspaces scales as $(1/\sigma^2)^{\frac{r}{2}}$.

For both linear and affine subspace classification, the lower bounds on performance are realized by any feature matrix having $M$ orthonormal rows in $\mathbb{R}^N$. Therefore, for regimes in which the bounds are tight, the asymptotic performance as characterized by the classification capacity and DDT is invariant to rotations of the linear features.

In Section \ref{sect:experiments}, we evaluate our claims empirically. We first examine the error performance of classifiers over randomly-drawn linear subspaces, focusing on the regimes in which the upper and lower bounds disagree.
Then, we test the correspondence of our theoretical results to a practical face recognition application. Using standard classification algorithms against public datasets, we observe error performance that agrees with our predictions to within a reasonable tolerance.

\subsection{Prior Work}
The statistics and machine learning literature contains a large body of work on feature extraction or supervised dimensionality reduction. In addition to the venerable linear discriminant analysis and principal component analysis, which depend only on the second-order statistics of the data, linear techniques based on 
higher-order statistics were developed in\cite{chen:ICML12,erdogmus:VLSI04,hild:PAMI06,kaski:ICML03,liu:PAMI12,nenadic:PAMI07,tao:PAMI09,torkkola:NIPS01,torkkola:JMLR03}. Owing to Fano's inequality, the algorithm of \cite{chen:ICML12} chooses linear features having maximal the mutual information with the classes, whereas \cite{hild:PAMI06,torkkola:NIPS01,torkkola:JMLR03} employ approximations to the mutual information based on R\'{e}nyi entropy. In \cite{qiu:arxiv13} linear features are chosen for subspace classification according to a nuclear-norm optimization problem, and in \cite{hamm:ICML08} an LDA-inspired Grassmann discriminant analysis is proposed. Finally, nonlinear dimensionality reduction techniques have recently become popular \cite{tenebaum:science00,wang:ICML2014}.

In the signal processing literature, information-theoretic limits on subspace classification arise under the framework of sparse support recovery. The set of all $k$-sparse vectors in $\mathbb{R}^N$ is a union of subspaces, and recovering the sparsity pattern is equivalent to finding the subspace in which the signal lies. A (data) deluge of recent works \cite{aeron:IT10,akcakaya:IT10,fletcher:IT09,rad:IT11,reeves:IT12,reeves:IT13,tang:IT10,tulino:IT13,wang:IT10,wainwright:IT09a,wainwright:IT09b,zhao:JMLR06,aksoylar:IT13} provides necessary and sufficient scaling laws on the triplet $(N, k, M)$, where $M$ is the number of compressive measurements taken, for recovery of a sparse signal. Different assumptions on the measurement matrices, decoders, error metrics, and sparsity regimes give rise to different scaling laws. While these works do provide fundamental limits on subspace classifier performance, sparse support recovery entails a specialization to the union of canonical subspaces, and the results presented in the preceding works do not bear directly on our study.

Reference \cite{calderbank:ICASSP12} considers compressed learning, i.e. learning directly in the compressive measurement domain rather than in the original data domain, showing that 
when data admit a sparse representation, low-dimensional feature extraction preserves the learnability and the separability of the data.
Along a similar vein, a recent work \cite{bandeira:14} considers the compressive classification of convex sets, proving limits on the number of measurements required to ensure that the projected sets remain separated.

A few works have focused on the classification of Gaussian mixtures, which is closely related to the linear and affine subspace classification considered herein. In \cite{reboredo:IT14} classifier performance is studied for a finite number of classes as a function of signal geometry; these results prefigure the DDT results presented in the sequel. In \cite{renna:IT13}, the number of measurements required to classify and reconstruct a signal drawn from a Gaussian mixture is characterized.

Researchers have also studied information-theoretic limits on other classification problems. The authors of \cite{santhanam:ISIT08} provide asymptotic limits on the success of model selection of Markov random fields. The authors of \cite{acarhya:NIPS12} use results in universal source coding to prove general bounds on classifier performance. The authors of \cite{tuncel:ISIT10} study the limits of database recovery from low-dimensional features, characterizing an ``identification capacity'' which is analogous to the classification capacity studied in this paper. 

\subsection{Notation}
We let bold lowercase letters denote vectors and bold uppercase letters to denote matrices. We let $\mathbb{R}$ and $\mathbb{Z}$  denote the field of reals and integers, respectively. We let $\mathbf{I}$ and $\mathbf{0}$ denote the identity matrix and the all-zeros matrix, respectively, indicating the size of the matrix in a subscript when necessary. We let $\norm{\cdot}$ denote the Euclidean norm; when applied to a matrix it denotes the induced operator norm. We let $E[\cdot]$  denote the expectation, indicating the distribution over which the expectation is taken by a subscript when necessary. We let $\lfloor \cdot \rfloor$ and $[\cdot]^+$ denote the floor function and the positive part of a number, respectively. We let $\mathrm{eig}(\cdot)$ denote the vector of eigenvalues of a square matrix. We let $\stackrel{d}{=}$  denote equality in distribution. We let $\mathcal{N}(\mu,\Sigma)$  denote a Gaussian distribution with mean $\mu$ and covariance matrix $\Sigma$. We let $\mathcal{W}_M(N,\mathbf{V})$ denote the $M \times M$ Wishart distribution with degrees of freedom $N$ and shape matrix $\mathbf{V}$.

\section{Preliminaries}\label{sect:preliminaries}

\subsection{Problem Definition}
We consider the statistical classification problem, in which the signal of interest $\mathbf{x} \in \mathbb{R}^N$ is distributed according to one of $L$ class-conditional densities $p_l(\mathbf{x})$, each of which is known to the classifier. The classifier observes noisy linear projections of $\mathbf{x}$, from which it attempts to determine the class-conditional density from which $\mathbf{x}$ was drawn. These projections, denoted by $\mathbf{y} \in \mathbb{R}^M$, are related to the signal $\mathbf{x} \in \mathbb{R}^N$ as follows:
\begin{equation}\label{eqn:signal.model}
	\mathbf{y} = \Phi\mathbf{x} + \mathbf{z},
\end{equation}
where $\Phi \in \mathbb{R}^{M \times N}$ is a matrix describing the linear features, and $\mathbf{z} \in \mathbb{R}^M$ is white Gaussian noise with mean zero and per-component variance $\sigma^2$ for some $\sigma^2 > 0$. We suppose $M \leq N$, and we constrain $\norm{\Phi} \leq 1$.
The noise $\mathbf{z}$ describes the deviation between the postulated subspace model and the true signal of interest.\footnote{Equivalently, we could remove the additive noise $\mathbf{z}$ and add a $\sigma^2\mathbf{I}$ term to the covariance matrix of each class-conditional density.} Signals will not lie perfectly on the linear or affine subspaces, so we suppose that the projected signal lies approximately within a ball centered at the specified subspaces and having radius $\sqrt{M}\sigma$.

To model the classification of linear and affine subspaces, we impose structure on the class-conditional densities $p_l(\mathbf{x})$. In particular, we suppose that the class conditional densities are Gaussian with low-rank covariances that correspond to the subspaces. In the case of linear subspaces, these Gaussians have zero mean. In the case of affine subspaces, which are simply translations of linear subspaces, the Gaussians have nonzero means.

We therefore define two sets. For the classification of $k$-dimensional linear subspaces of $\mathbb{R}^N$, the class-conditional densities $p_l(\mathbf{x})$ belong to\footnote{We will drop the subscripts $\mathrm{linear}$ and $\mathrm{affine}$ throughout when discussing classification generally rather than particularizing to linear or affine subspaces.}
\begin{equation}
	\mathcal{Q}_\mathrm{linear}(N,k) = \{ \mathcal{N}(\mathbf{0},\mathbf{U}\mathbf{U}^T) : \mathbf{U} \in \mathbb{R}^{N \times k} \}.
\end{equation}
In other words, each class-conditional density is a Gaussian supported on the $k$-dimensional subspace spanned by the columns of $\mathbf{U}$. Similarly, for the classification of $k$-dimensional affine spaces, the class-conditional densities belong to
\begin{equation}
		\mathcal{Q}_\mathrm{affine}(N,k) = \{\mathcal{N}(\mu,\mathbf{U}\mathbf{U}^T) : \mu \in \mathbb{R}^{N}, \mathbf{U} \in \mathbb{R}^{N \times k} \}.
\end{equation}
That is, the class-conditional densities are supported on a $k$-dimensional subspace as before, but here they are translated by a non-zero vector $\mu$.

We parameterize the sets $\mathcal{Q}_\mathrm{linear}(N,k)$ and $\mathcal{Q}_\mathrm{affine}(N,k)$ by the following two sets
\begin{align}
	\mathcal{A}_\mathrm{linear}(N,k) &= \mathbb{R}^{N \times k} \\
	\mathcal{A}_\mathrm{affine}(N,k) &= \mathbb{R}^N \times \mathbb{R}^{N \times k}.
\end{align}
Clearly $\mathcal{A}_\mathrm{linear}(N,k)$ and $\mathcal{A}_\mathrm{affine}(N,k)$ are isomorphic to $\mathcal{Q}_\mathrm{linear}(N,k)$ and $\mathcal{Q}_\mathrm{affine}(N,k)$, respectively. We can represent a linear or affine subspace classification problem by a tuple $\mathbf{a} = (a_1, \cdots, a_L) \in \mathcal{A}^L(N,k)$, where $\mathcal{A}^L$ is the $L$-fold Cartesian product of $\mathcal{A}(N,k)$. The tuple $\mathbf{a}$ encodes the $L$ covariances and, when appropriate, the $L$ means corresponding to the subspaces to be classified. Let
$ p(\mathbf{x} ; a_l) = p_l(\mathbf{x})$, for $1 \leq l \leq L$, denote the class-conditional densities parameterized by $\mathbf{a} \in \mathcal{A}(N,k)$.

Let $\hat{l} = f(\mathbf{y})$ denote the classifier output, where $f$ is a mapping from $\mathbb{R}^M$ to $\{1,\cdots,L\}$. Then, for a classification problem described by the tuple $\mathbf{a}$, define the average misclassification probability:
\begin{equation}\label{eqn:misclassification.probability}
	\mathrm{P}_e(\mathbf{a}) = \min_{\norm{\Phi} \leq 1} \frac{1}{L}\sum_{i=1}^L \mathrm{Pr}(\hat{l} \neq l | \mathbf{x} \sim p(\mathbf{x} ; a_l)),
\end{equation}
where each term in the sum is the misclassification probability when $\mathbf{x}$ is drawn according to $p(\mathbf{x} ; a_l)$. Observe that we define $\mathrm{P}_e(\mathbf{a})$ in terms of the best feature matrix $\Phi$. Therefore, in proving our results we will characterize the feature matrix that achieves optimal classifier performance in the asymptote.

The focus of this paper is the analysis of $\mathrm{P}_e$ in two asymptotic regimes: (i) as the signal dimensions $N,M,k$ go to infinity, and (ii) as the noise power $\sigma^2$ goes to zero. In the first case, we derive conditions for which the probability of error decays to zero, except for a set of vanishing probability over $\mathcal{A}^L$. In the second case, we derive scaling laws on the probability of error, {\em averaged} over the possible choices of $\mathbf{a} \in \mathcal{A}^L$. To this end, define the following probability distributions over the parameter sets $\mathcal{A}_\mathrm{linear}(N,k)$ and $\mathcal{A}_\mathrm{affine}(N,k)$:
\begin{align}
	p_\mathrm{linear}(a) &= \prod_{i=1}^N\prod_{j=1}^k \mathcal{N}(u_{ij}; 0, 1/k) \\
	p_\mathrm{affine}(a) &= \prod_{i=1}^N\prod_{j=1}^k \mathcal{N}(u_{ij}; 0, 1/k) \cdot \prod_{l=1}^N \mathcal{N}(\mu_i; 0, 1),
\end{align}
where $u_{ij}$ is the $(i,j)$th element of the matrix $\mathbf{U}$ and $\mu_i$ is the $i$th element of the vector $\mu$. These distributions define a measure over the sets of class-conditional densities. In other words, in computing probabilities we suppose that the elements of the matrix $\mathbf{U}$ and the mean vector $\mu$ are standard i.i.d. Gaussian.

For both $p_\mathrm{linear}$ and $p_\mathrm{affine}$, the distribution is supported over the entire parameter space, is invariant to rotations, and yields finite expected signal energy. Specifically, the distribution over the bases $\mathbf{U}$ is isotropic, which means that the linear subspaces are drawn uniformly from the Grassmann manifold. Therefore, our analysis characterizes classifier performance when ``nature'' presents us with subspaces without favoring a particular region of the Grassmann; we contend that this assumption is reasonable. Furthermore, while changes to the distributions $p_\mathrm{linear}$ and $p_\mathrm{affine}$ {\em will} change the classification capacity and DDT in general, the coarse behavior is robust to variations. In particular, one can recover our proofs subject to straightforward constraints on the eigenvalue distribution of $\mathbf{U}^T\mathbf{U}$, showing bounds on the classification capacity that differ at most by a $O(1)$ term and DDT bounds that agree exactly.

Next, we define the classification capacity and the diversity-discrimination tradeoff.

\subsection{Classification Capacity}
The classification capacity characterizes fundamental performance limits as the signal dimensions approach infinity. We derive bounds on how fast the number of subspaces can grow, as a function of $N$, $M$, and $k$, while ensuring the misclassification probability decays to zero almost surely.

By analogy with the sequence of codebooks defined for the Shannon capacity, we characterize the classification capacity in terms of a sequence of classification problems indexed by $M$. We let the number of features $M$ grow to infinity, and we let the dimensions $N$ and $k$ scale linearly with $M$ as follows:
\begin{equation}
	N(M) = \lfloor \nu M \rfloor, k(M) = \lfloor \kappa M \rfloor,
\end{equation}
for $\nu \geq 1$ and $0 < \kappa < 1$. We also let the number of subspaces $L$ scale exponentially in $M$ as follows:
\begin{equation}
	L(M) = \lfloor 2^{\rho M} \rfloor,
\end{equation}
for some $\rho \geq 0$. By analogy with communications theory, the quantity $\rho$ can be interpreted as the ``rate'' of the sequence of class alphabets, or the average number of bits discerned per feature if  classification succeeds. Indeed, in the sequel we refer to $\rho$ a the {\em classification rate}.

\begin{definition}
	Fix the dimension ratios $\nu$ and $\kappa$ and the classification rate $\rho$. Then, define the set of classification problems for which the probability of classification error exceeds an arbitrary small constant $\epsilon > 0$:
	\begin{equation}
		\mathcal{E}(M) = \{\mathbf{a} \in \mathcal{A}^{L(M)}(N(M),k(M)) : \mathrm{P}_e(\mathbf{a}) > \epsilon\}.
	\end{equation}
	Then, we say that $\rho$ is {\em achievable} provided
	\begin{equation}
		\lim_{M \to \infty} \int_{\mathcal{E}(M)} \prod_{i=1}^{L(M)} p(a_i) d\mathbf{a} = 0,
	\end{equation}
	for any fixed $\epsilon > 0$. 
\end{definition}

Observe that a classification rate $\rho$ is achievable if
\begin{equation}
	\lim_{M \to \infty} E[\mathrm{P}_e(\mathbf{a})] = \lim_{M \to \infty} \int_{\mathcal{A}^{L(M)}(N(M),k(M))} \mathrm{P}_e(\mathbf{a}) \prod_{i=1}^{L(M)} p(a_i) d\mathbf{a} = 0.
\end{equation}
This observation follows by contradiction: If there is a subset of $\mathcal{A}^L(N,k)$ having non-trivial probability for which the misclassification probability remains bounded away from zero, the expected error also remains bounded away from zero.
\begin{definition}
	Fix the dimension ratios $\nu$ and $\kappa$. The {\em classification capacity}, denoted by $C_{\mathrm{linear}}(\nu,\kappa)$ and $C_{\mathrm{affine}}(\nu,\kappa)$ for linear and affine space classification, respectively, is the supremum over achievable classification rates $\rho$.
\end{definition}

In other words, if the classification rate is smaller than $C(\nu,\kappa)$, then the probability of classification error approaches zero almost surely over the set of subspace classification problems. Otherwise, the error probability remains bounded away from zero for a non-trivial subset of $\mathcal{A}^L(N,k)$.

Although the classification capacity is defined to characterize classifier behavior when $N$ and $k$ scale linearly in $M$ and $L$ scales exponentially in $M$, it also captures other regimes automatically. For example, if $k$ scales sub-linearly in $M$, the asymptotic behavior is the same as if $\kappa \to 0$. Similarly, if $L$ scales sub-exponentially in $M$, the misclassification probability decays to zero whenever the classification capacity is nonzero, and if $L$ scales super-exponentially the misclassification capacity remains bounded from zero whenever the classification capacity is finite. In view of Theorems \ref{thm:zero.mean.capacity} and \ref{thm:nonzero.mean.capacity}, this implies that whenever $\kappa >0$ and the number of subspaces grows polynomially in $M$, the misclassification probability goes to zero. Because we are dealing with subspaces, it is impossible to have $N$ scale sub-linearly or $k$ scale super-linearly in $M$. However, at least one regime remains unspecified by our analysis: If $N$ scales super-linearly in $M$, classifier behavior is unclear.

We can bound the classification capacity via the mutual information between the vector $\mathbf{a} \in \mathcal{A}$ and the feature vector $\mathbf{y}$.
\begin{lemma}\label{thm:fano}
	The classification capacity satisfies
	\begin{equation}\label{eqn:capacity.upper.bound}
		C \leq \lim_{M \to \infty} \max_{\norm{\Phi} \leq 1} \frac{I(\mathbf{a}; \mathbf{y})}{M},
	\end{equation}
	where the mutual information is computed with respect to $p_\mathrm{linear}(a)$ or $p_\mathrm{affine}(a)$ as appropriate.
\end{lemma}
\begin{IEEEproof}
	The proof follows from Fano's inequality. By the standard arguments (e.g. from \cite{cover:06}), we obtain
	\begin{equation*}
		P_e(\mathrm{a}) \geq  1 - \frac{ \max_{\norm{\Phi} \leq 1} I(\mathbf{a}; \mathbf{y}) - 1}{M \rho},
	\end{equation*}
	which is bounded away from zero when $\rho$ exceeds the RHS of (\ref{eqn:capacity.upper.bound}).
\end{IEEEproof}

Observe that we have proven more than just an upper bound on the capacity. When $\rho$ exceeds RHS of (\ref{eqn:capacity.upper.bound}), not only is there a non-trivial set for which the error probability remains positive, but that set is also all of $\mathcal{A}^L(N,k)$. If the upper bound of Lemma \ref{thm:fano} is tight, then the mutual information characterizes a sharp phase transition in the error probability. If the number of subspaces grows sufficiently slowly, the probability of error vanishes almost everywhere; otherwise, is bounded away from zero everywhere.

However, it is not clear whether Lemma \ref{thm:fano} is tight. If the ``channel'' between subspaces and features is {\em information stable}---meaning roughly that the normalized information density converges on the normalized mutual information---then the mutual information completely characterizes the classification capacity and (\ref{eqn:capacity.upper.bound}) holds with equality \cite{dobrushin:AMS63}. Alternatively, applying the results of \cite{verdu:IT94}, one can express the classification capacity directly in terms of the information density. Analysis of the information density is difficult, however, as is the verification of information stability, so Lemma \ref{thm:fano} remains an upper bound only.

To prove lower bounds on the classification capacity, we analyze directly the misclassification probability. Our main tool is the Bhattacharyya bound on the pairwise misclassification probability\cite{bhattacharyya:IJS46,kailath:TC67}, which we state here for Gaussian distributions.
\begin{lemma}\label{thm:bhattacharyya}
	Suppose we observe a signal that is distributed according to $\mathcal{N}(\mu_1,\Sigma_1)$ or $\mathcal{N}(\mu_2,\Sigma_2)$ with equal prior probability. Define
	\begin{equation}
		B =  \frac{1}{2} \ln \left(\frac{|\frac{\Sigma_1 + \Sigma_2}{2}|}{|\Sigma_1|^{\frac{1}{2}}|\Sigma_2|^{\frac{1}{2}}}\right) + \frac{1}{8}(\mu_1-\mu_2)\left[\frac{\Sigma_1 + \Sigma_2}{2}\right]^{-1}(\mu_1-\mu_2).
	\end{equation}
	Then, supposing maximum likelihood classification, the misclassification probability is bounded by
	\begin{equation}
		\mathrm{P}_e((\mu_1,\Sigma_1, \mu_2,\Sigma_2)) \leq \frac{1}{2} \exp(-B).
	\end{equation}
\end{lemma}
In \cite{kailath:TC67} it is also observed that the Bhattacharyya bound is exponentially tight in the sense that, if the pairwise error decays to zero, it approaches $c \cdot \exp(-B)$ for some constant $c$. A consequence of this observation, which we will see in Section \ref{sect:ddt}, is that the Bhattacharyya bound predicts the maximum diversity gain for both linear and affine subspace classifiers.

\subsection{Diversity-Discrimination Tradeoff}
The {\em diversity-multiplexing tradeoff} (DMT) was introduced in the context of wireless communications to characterize the high-SNR performance of fading vector channels. It was shown in \cite{zheng:IT03} that the spatial flexibility provided by multiple antennas can simultaneously increase the achievable rate and decrease the probability of error, but only according to a tradeoff that is precisely characterized at high SNR. We define a similar characterization in the context of classification, called the {\em diversity-discrimination tradeoff} (DDT), which captures the relationship between the increase of discernible subspaces and the decay of misclassification probability as the noise power approaches zero.

For the DDT, we keep $N$, $M$, and $k$ fixed, but we let the number of subspaces scale in the noise power as follows:
\begin{equation}
	L(\sigma^2) = \lfloor (1/\sigma^2)^{\frac{r}{2}} \rfloor,
\end{equation}
for some $r \geq 0$, which we call the {\em discrimination gain}. We define the DDT in terms of the misclassification probability averaged over the ensemble of classification problems, which we denote by
\begin{equation}
	\bar{\mathrm{P}}_e(\sigma^2,r) = E[\mathrm{P}_e(\mathbf{a})] = \int_{\mathcal{A}(N,k)} \mathrm{P}_e(\mathbf{a}) \prod_{l=1}^{L(\sigma^2)}p(a_l) d\mathbf{a},
\end{equation}
where here we express the probability of error as a function of the discrimination gain $r$ and the noise power $\sigma^2$. Specifically, the diversity-discrimination tradeoff is defined as the following function:
\begin{equation}
	d(r) = \lim_{\sigma^2 \to 0} -\frac{\log \bar{\mathrm{P}}_e(\sigma^2,r)}{\frac{1}{2}\log (1/\sigma^2)}.
\end{equation}
We refer to $d(r)$ as the {\em diversity gain} for discrimination gain $r$. In other words, when the number of subspaces increases as $(1/\sigma^2)^{r/2}$, the probability of error decays as $(1/\sigma^2)^{-d(r)/2} + o(\log(\sigma^2))$. In the sequel we refer to $d_\mathrm{linear}(r)$ and $d_\mathrm{affine}(r)$ as appropriate.

By contrast to the classification capacity, where we characterize phase transitions in the error probability that hold almost surely, for the DDT we specify scaling laws in the error probability that hold on the average over $\mathcal{A}$. Rather than specifying {\em if} the probability of error decays to zero, the DDT specifies {\em how quickly} it decays. In the former case, it is straightforward to define the failure event and show that it has vanishing probability. In the latter case, it is unclear how to define such a failure event, so we state only an average-case result.

As with the classification capacity, we can derive bounds on the DDT from the mutual information.
\begin{lemma}\label{thm:fano.ddt}
	Fix $N$, $M$, and $k$. Then, $d(r)=0$ whenever
	\begin{equation}
		r \geq \lim_{\sigma^2 \to 0} \max_{\Phi, \norm{\Phi} \leq 1} \frac{I(\mathbf{a}; \mathbf{y})}{\frac{1}{2}\log (1/\sigma^2)},
	\end{equation}
	where again the mutual information is calculated with respect to $p_\mathrm{linear}(a)$ or $p_\mathrm{affine}(a)$ as appropriate.
\end{lemma}
\begin{IEEEproof}
	Again we invoke Fano's inequality. Whenever $r$ is as large as the specified quantity, the probability of error is bounded away from zero, and the diversity gain is zero by definition.
\end{IEEEproof}


\section{Classification Capacity}\label{sect:capacity}
Here we characterize the classification capacities $C_\mathrm{linear}(\nu,\kappa)$ and $C_\mathrm{affine}(\nu,\kappa)$. We prove upper bounds that show that, for both linear and affine subspace classification, the probability of error remains bounded away from zero almost surely whenever the number of subspaces scales faster than $(1/\sigma^2)^{\frac{M-k}{2}}$. For linear spaces, we prove a lower bound which matches the upper bound to within an $O(1)$ term for $\kappa \geq 1/2$; otherwise the bounds disagree. For affine spaces, we prove a lower bound which is tight to within an $O(1)$ term for all $\kappa$. This suggests the somewhat surprising conclusion that, at least for $\kappa \geq 1/2$, translating subspaces by nonzero vectors does not substantially increase the number of subspaces a classifier can discriminate. Whether this conclusion extends to $\kappa < 1/2$ depends on the tightness of the upper bound. However, as we will see in Section \ref{sect:ddt}, affine subspaces are easier to discriminate in the sense that the misclassification probability decays faster as $\sigma^2 \to 0$.

\subsection{Linear Subspaces}\label{sect:zero.mean.capacity}
First, we bound on $C_\mathrm{linear}(\nu,\kappa)$.
\begin{theorem}\label{thm:zero.mean.capacity}
For linear subspace classification, the classification capacity is bounded by
	\begin{multline}
		 \frac{\min\{\kappa,1-\kappa\}}{2}\log_2\left(1+ \frac{(\sqrt{1/(2\kappa)}-1)^2}{\sigma^2}\right) - \frac{\kappa}{2} \leq C_\mathrm{linear}(\nu,\kappa) \leq \\
		 \frac{1-\kappa}{2}\log_2\left(\frac{1}{\sigma^2}\right) + \frac{1}{2}\log_2(1+\sigma^2) - \frac{\kappa}{2}\log_2((\sqrt{1/\kappa}-1)^2 + \sigma^2).
	\end{multline}
\end{theorem}
\begin{IEEEproof}
We first prove the upper bound by estimating the mutual information between the subspaces and the features and invoking Lemma \ref{thm:fano}. Then, we prove the lower bound by invoking Lemma \ref{thm:bhattacharyya} and applying the union bound.

{\bf Upper Bound:}
To bound the mutual information $I(\mathbf{a}; \mathbf{y}) = I(\mathbf{U};\mathbf{y})$, we first characterize the optimum choice of $\Phi$. Following the argument in \cite[Theorem 2]{chen:ICML12}, we compute the gradient of the mutual information with respect to the singular values of $\Phi$. Writing the singular value decomposition as $\Phi = \mathbf{W}_\Phi\Lambda_\Phi\mathbf{V}_\Phi^T$, the gradient is
\begin{equation}\label{eqn:mi.gradient}
	\nabla_{\Lambda_\Phi} I(\mathbf{U};\mathbf{y}) = \Lambda_\Phi \mathbf{V}_\Phi^T E\left[ \int p(\mathbf{y}|\mathbf{U})(\mathbf{m} - \mathbf{m}_\mathbf{U}) (\mathbf{m} - \mathbf{m}_\mathbf{U})^T d\mathbf{y}\right] \mathbf{V}_\Phi,
\end{equation}
where
\begin{equation*}
	\mathbf{m} = \int \mathbf{x} p(\mathbf{x} | \mathbf{y}) d\mathbf{x}
\end{equation*}
is the mean with respect to the posterior distribution, and
\begin{equation*}
	\mathbf{m}_\mathbf{U} = \int \mathbf{x} p(\mathbf{x} | \mathbf{y},\mathbf{U}) d\mathbf{x}
\end{equation*}
is the mean with respect to the conditional posterior. Observe from (\ref{eqn:mi.gradient}) that the diagonal elements of the gradient are non-negative, which implies that the mutual information is non-decreasing with the singular values of $\Phi$. Because we constrain $\norm{\Phi} \leq 1$, it follows that the singular values of the optimal $\Phi$ are identically unity.

Assuming this condition on $\Phi$, we bound the mutual information. By definition,
\begin{equation*}
	I(\mathbf{U};\mathbf{y}) = h(\mathbf{y}) - h(\mathbf{y} |  \mathbf{U}).
\end{equation*}
To bound the conditional entropy, observe that the conditional distribution of $\mathbf{y}$ is
\begin{equation*}
	p(\mathbf{y} | \mathbf{U}) = \mathcal{N}(0, \Phi\mathbf{U}\mathbf{U}^T\Phi^T +\sigma^2 \cdot \mathbf{I}).
\end{equation*}
Let $\lambda_i$ denote the $i$th ordered eigenvalue of $\mathbf{U}^T\Phi^T\Phi\mathbf{U}$. Then, the conditional entropy is
\begin{align*}
	h(\mathbf{y} | \mathbf{U}) &= \sum_{i=1}^k\frac{1}{2}E[\log_2(2\pi e(\lambda_i + \sigma^2))] + \frac{M-k}{2}\log_2(2\pi e \sigma^2) \\
	&\geq \frac{k}{2}E\left[\log_2 (\lambda_k + \sigma^2) \right] + \frac{M-k}{2}\log_2(\sigma^2) + \frac{M}{2}\log_2(2\pi e), \label{eqn:conditional.entropy.determinant}
\end{align*}
where the expectation is with respect to $\mathbf{U}$, and where the inequality is trivially obtained by substituting the smallest positive eigenvalue $\lambda_k$. We next bound this eigenvalue. Because $\Phi$ has singular values identically equal to unity, and because $\mathbf{U}$ has i.i.d. Gaussian entries with variance $1/k$, the matrix $\Phi \mathbf{U} \in \mathbb{R}^{M \times k}$ also has i.i.d. Gaussian entries with variance $1/k$. Therefore,
\begin{equation*}
	\mathbf{U}^T\Phi^T \Phi \mathbf{U} \stackrel{d}{=} \frac{1}{k}\mathbf{W},
\end{equation*}
where $\mathbf{W} \sim \mathcal{W}_k(\mathbf{I},M)$.
In \cite[Theorem 1]{silverstein:AP85} it is shown that the smallest eigenvalue of $1/M \cdot \mathbf{W}$ converges to $(1-\sqrt{\kappa})^2$ almost surely as $M \to \infty$. Therefore, the minimum eigenvalue of $\mathbf{U}^T\Phi^T \Phi \mathbf{U}$, which is equal to $\lambda_k$, converges on $(\sqrt{1/\kappa}-1)^2$ almost surely. We can therefore bound the conditional mutual information by
\begin{equation}\label{eqn:zero.mean.conditional.entropy}
	h(\mathbf{y} | \mathbf{U}) \geq \frac{k}{2}\log_2((\sqrt{1/\kappa}-1)^2 + \epsilon(M) + \sigma^2) + \frac{M-k}{2}\log_2(\sigma^2) + \frac{M}{2}\log_2(2\pi e) ,
\end{equation}
where $\epsilon(M) \to 0$ as $M \to \infty$.

We turn next to the differential entropy of $\mathbf{y}$. We first compute the expected covariance, which is
\begin{align*}
	E[\mathbf{y}\mathbf{y}^T] &= E_\mathbf{U}[\Phi\mathbf{U} \mathbf{U}^T \Phi + \sigma^2\mathbf{I}] \\
	&= (1+\sigma^2)\mathbf{I}. 
\end{align*}
Noting that the differential entropy for a fixed covariance is maximized by the multivariate Gaussian distribution, we obtain
\begin{equation}\label{eqn:zero.mean.marginal.entropy}
	h(\mathbf{y}) \leq \frac{M}{2}\log_2(2\pi e(1 + \sigma^2)).
\end{equation}
Combining terms and letting $M \to \infty$, we finally obtain
\begin{equation}\label{eqn:normalized.mutual.information}
	\lim_{M \to \infty} \max_{\Phi} \frac{I(\mathbf{y}; \mathbf{U})}{M} \leq \frac{1-\kappa}{2}\log\left(\frac{1}{\sigma^2}\right) + \frac{1}{2}\log_2(1+\sigma^2) - \frac{\kappa}{2}\log_2((\sqrt{1/\kappa}-1)^2 + \sigma^2).
\end{equation}
Applying Lemma \ref{thm:fano} to (\ref{eqn:normalized.mutual.information}), we obtain the upper bound.

{\bf Lower Bound:}
Choose $\Phi \in \mathbb{R}^{M \times N}$ to be any matrix with orthonormal rows. Observe that while this choice maximizes the mutual information, it does not minimize the probability of error in general. Applying the Bhattacharyya bound from Lemma \ref{thm:bhattacharyya}, the probability of a pairwise error between two subspaces $i$ and $j$ is bounded by
\begin{equation*}
	\mathrm{P}_e(\mathbf{U}_i,\mathbf{U}_j) \leq \frac{1}{2} \cdot \left(\frac{|\frac{\Phi \mathbf{U}_i \mathbf{U}_i^T\Phi^T + \Phi \mathbf{U}_j \mathbf{U}_j^T\Phi^T + 2\sigma^2\mathbf{I}}{2}|}{|\Phi \mathbf{U}_i \mathbf{U}_i^T\Phi + \sigma^2 \mathbf{I}|^{\frac{1}{2}}|\Phi \mathbf{U}_j \mathbf{U}_j^T \Phi^T + \sigma^2 \mathbf{I}|^{\frac{1}{2}}}\right)^{-\frac{1}{2}}.
\end{equation*}
With probability one, the matrices $\Phi \mathbf{U}_i \mathbf{U}_i^T\Phi^T$ and $\Phi \mathbf{U}_j \mathbf{U}_j^T\Phi^T$ have rank $k$, and the matrix $(\Phi \mathbf{U}_i \mathbf{U}_i^T + \Phi \mathbf{U}_j \mathbf{U}_j^T\Phi)/2$ has rank $\min\{M,2k\}$. Let $\lambda_{i_{l}}$ and $\lambda_{j_{l}}$ denote the nonzero eigenvalues of $\Phi \mathbf{U}_i \mathbf{U}_i^T$ and $\Phi \mathbf{U}_j \mathbf{U}_j^T$, respectively, and let $\lambda_{ij_l}$ denote the nonzero eigenvalues of the latter matrix. Then, we can write the pairwise bound as
\begin{align*}
	\mathrm{P}_e(\mathbf{U}_i,\mathbf{U}_j) &\leq \frac{1}{2} \left(\frac{(\sigma^2)^{M-\min\{M,2k\}}\prod_{l=1}^{\min \left(2k,M\right)} \left(\lambda_{ij_l} + \sigma^2\right)  }{\sqrt{(\sigma^2)^{M-k}\prod_{l=1}^k \left(\lambda_{i_l} + \sigma^2\right) \cdot (\sigma^2)^{M-k}\prod_{l=1}^k \left(\lambda_{j_l} + \sigma^2\right)}}\right)^{-\frac{1}{2}} \\
	&= \frac{1}{2} \cdot \left(\frac{1}{\sigma^2}\right)^{-\frac{\min\{M-k,k\}}{2}} \cdot
	\left(\frac{\prod_{l=1}^{\min \left(2k,M\right)}  \left(\lambda_{ij_l} + \sigma^2 \right)}
	{\sqrt{\prod_{l=1}^{k} (\lambda_{i_l} + \sigma^2 ) \cdot
	\prod_{l=1}^{k} (\lambda_{j_l} + \sigma^2 )}}\right)^{-\frac{1}{2}}.
\end{align*}
By construction,
\begin{equation*}
	\Phi \mathbf{U}_i \mathbf{U}_i^T\Phi^T + \Phi \mathbf{U}_j \mathbf{U}_j^T\Phi^T \geq \Phi \mathbf{U}_i \mathbf{U}_i^T\Phi^T, \Phi \mathbf{U}_j \mathbf{U}_j^T\Phi^T.
\end{equation*}
By Weyl's monotonicity theorem (see, e.g., \cite{horn:12}), $2 \lambda_{ij_l} \geq \lambda_{i_l}$ and $2 \lambda_{ij_l} \geq \lambda_{j_l}$for every $1 \leq l \leq k$. Therefore,
\begin{equation*}
	\prod_{l=1}^{k} 2 \left(\lambda_{ij_l} + \sigma^2\right) \geq \sqrt{\prod_{l=1}^{k} (\lambda_{i_l} + \sigma^2 ) \cdot
	\prod_{l=1}^{k} (\lambda_{j_l} + \sigma^2 )},
\end{equation*}
from which it follows that
\begin{align*}
	\mathrm{P}_e(\mathbf{U}_i,\mathbf{U}_j) &\leq \frac{1}{2} \cdot \left(\frac{1}{\sigma^2}\right)^{-\frac{\min\{M-k,k\}}{2}} \cdot 2^{\frac{k}{2}} \cdot \left(\prod_{l=k+1}^{\min \left(2k,M\right)}  \left(\lambda_{ij_l} + \sigma^2 \right)\right)^{-\frac{1}{2}} \\
	&\leq \frac{1}{2} \cdot \left(\frac{1}{\sigma^2}\right)^{-\frac{\min\{M-k,k\}}{2}} \cdot 2^{\frac{k}{2}} \cdot \left(\lambda_{ij_{\min\{2k,M\}}} + \sigma^2\right)^{-\frac{\min\{M-k,k \}}{2}} \\
	&= 2^{\frac{k-2}{2}} \cdot \left(1 + \frac{\lambda_{ij_{\min\{2k,M\}}}}{\sigma^2}\right)^{-\frac{\min\{M-k,k\}}{2}}.
\end{align*}

Next, we bound the eigenvalue $\lambda_{{ij}_{\min\{2k,M\}}}$. Because each matrix $\mathbf{U}_i$ has i.i.d. Gaussian entries with zero mean and variance $1/k$, so too does each matrix $\Phi \mathbf{U}_l \in \mathbb{R}^{M \times k}$. Furthermore, observe that
\begin{equation*}
	\Phi\mathbf{U}_i \mathbf{U}_i^T\Phi^T + \Phi\mathbf{U}_j \mathbf{U}_j^T\Phi^T =
	\begin{bmatrix}
		\Phi\mathbf{U}_i & \Phi\mathbf{U}_j
	\end{bmatrix} \cdot
	\begin{bmatrix}
		(\Phi\mathbf{U}_i)^T \\ (\Phi\mathbf{U}_j)^T
	\end{bmatrix}.
\end{equation*}
Therefore, the nonzero eigenvalues of $\Phi\mathbf{U}_i \mathbf{U}_i^T\Phi^T + \Phi\mathbf{U}_j \mathbf{U}_j^T\Phi^T$ are those of a scaled Wishart matrix. Specifically, if $2k < M$, the eigenvalues are those of $1/k \cdot \mathbf{W}_1$, where $\mathbf{W}_1 \sim \mathcal{W}_{2k}(M,\mathbf{I})$. If $M \geq 2k$, the eigenvalues are those of $1/k \cdot \mathbf{W}_2$, where $\mathbf{W}_2 \sim \mathcal{W}_M(2k,\mathbf{I})$.
By \cite[Theorem 1]{silverstein:AP85}, the minimum eigenvalue in either case converges on $(\sqrt{1/\kappa} - \sqrt{2})^2$ almost surely. Therefore, $\lambda_{{ij}_{\min\{2k,M\}}}$ converges on $(\sqrt{1/(2\kappa)}-1)^2$ almost surely, and we obtain
\begin{equation}\label{eqn:zero.mean.pairwise.error}
	\mathrm{P}_e(\mathbf{U}_i,\mathbf{U}_j) \leq 2^{\frac{k-2}{2}} \cdot \left(1 + \frac{(\sqrt{1/(2\kappa)}-1)^2 +\epsilon(M)}{\sigma^2}\right)^{-\frac{\min\{M-k,k\}}{2}},
\end{equation}
where $\epsilon(M) \to 0$ almost surely as $M \to \infty$. {\em A fortiori}, the bound in (\ref{eqn:zero.mean.pairwise.error}) is also a bound on the {\em expected} pairwise probability, with $\epsilon(M)$ independent for each $i,j$ pair.

Invoking the union bound over all $L(M)$ subspaces, we obtain
\begin{align*}
	E[\mathrm{P}_e(\mathbf{a})] &\leq \frac{1}{L(M)}\sum_{l=1}^{L(M)}\sum_{l^\prime \neq l} E[\mathrm{P}_e(\mathbf{U}_l,\mathbf{U}_{l^\prime})] \\
	&= (L(M) - 1) E[\mathrm{P}_e(\mathbf{U}_l,\mathbf{U}_{l^\prime})] \\
	&\leq 2^{\rho M} E[\mathrm{P}_e(\mathbf{U}_l,\mathbf{U}_{l^\prime})],
\end{align*}
where the second equality follows because each $\mathbf{U}_l$ is drawn independently. Taking the logarithm of both sides yields
\begin{equation}
	\log_2 (E[\mathrm{P}_e(\mathbf{a})]) \leq \rho M + \frac{k-2}{2} - \frac{\min\{M-k,k\}}{2}\log_2\left(1 + \frac{(\sqrt{1/(2\kappa)} - 1)^2 +\epsilon(M)}{\sigma^2}\right).
\end{equation}
Therefore, if
\begin{equation*}
	\rho < \frac{\min\{1-\kappa,\kappa \}}{2}\log_2\left(1 + \frac{(\sqrt{1/(2\kappa)} - 1)^2}{\sigma^2}\right) - \frac{\kappa}{2},
\end{equation*}
then $E[\mathrm{P}_e(\mathbf{a})]$ goes to zero as $M \to \infty$, and thus $\mathrm{P}_e(\mathbf{a})$ goes to zero almost surely, as was to be shown.
\end{IEEEproof}

When $\kappa \geq \frac{1}{2}$, the lower and upper bounds agree to within a $O(1)$ term; otherwise they are loose. Based on the numerical experiments presented in Section \ref{sect:experiments}, we conjecture that the upper bound is approximately tight, while the lower bound is loose.

\subsection{Affine Subspaces}
Next, we bound $C_\mathrm{affine}(\nu,\kappa)$.
\begin{theorem}\label{thm:nonzero.mean.capacity}
For affine subspace classification, the classification capacity satisfies
\begin{equation}
	C_\mathrm{affine}(\nu,\kappa) \leq \frac{1-\kappa}{2}\log_2\left(\frac{1}{\sigma^2}\right) + \frac{1}{2}\log_2(2+\sigma^2) - \frac{\kappa}{2}\log_2((\sqrt{1/\kappa}-1)^2 + \sigma^2),
\end{equation}
and
\begin{equation}
	C_\mathrm{affine}(\nu,\kappa) \geq 
	\begin{cases}
		\frac{1-\kappa}{2}\log_2\left(1+ \frac{\min\{(\sqrt{1/(2\kappa)}-1)^2,1/2\}}{\sigma^2}\right) - \frac{\kappa}{2} & \text{ for $\kappa < 1/2$} \\
		\frac{1-\kappa}{2}\log_2\left(1+ \frac{(\sqrt{1/(2\kappa)}-1)^2}{\sigma^2}\right) - \frac{\kappa}{2} & \text{ for $\kappa \geq 1/2$ }
	\end{cases}.
\end{equation}
\end{theorem}
\begin{IEEEproof}
As before, we prove the upper bound by bounding the mutual information, and the lower bound by direct analysis of the probability of error via the Bhattacharyya bound.

{\bf Upper Bound:}
To prove the upper bound, we expand the mutual information as
\begin{equation}
	I(\mathbf{a} ; \mathbf{y}) = I(\mathbf{U},\mu ; \mathbf{y}) = h(\mathbf{y}) - h(\mathbf{y} | \mathbf{U},\mu).
\end{equation}
As in the case of linear subspaces, the $\Phi$ that maximizes the mutual information has unit singular values. Furthermore, because the entropy of a Gaussian does not depend on the mean, $h(\mathbf{y} | \mathbf{U},\mu) = h(\mathbf{y} | \mathbf{U})$. Therefore, applying (\ref{eqn:zero.mean.conditional.entropy}), we obtain
\begin{equation}\label{eqn:affine.conditional.entropy}
	h(\mathbf{y} | \mathbf{U},\mu) \geq \frac{k}{2}\log_2((\sqrt{1/\kappa}-1)^2 + \epsilon(M) + \sigma^2) + \frac{M-k}{2}\log_2(\sigma^2) + \frac{M}{2}\log_2(2\pi e),
\end{equation}
where $\epsilon(M) \to 0$. Then, observing that
\begin{align*}
	E[\mathbf{y} \mathbf{y}^T] &= E_\mu[\Phi\mu \mu^T\Phi^T] + E_\mathbf{U}[\Phi\mathbf{U}\mathbf{U}^T\Phi^T] + \sigma^2 \mathbf{I} \\
	&= (2+\sigma^2)\mathbf{I},
\end{align*}
we conclude that
\begin{equation}\label{eqn:affine.marginal.entropy}
	h(\mathbf{y}) \leq \frac{M}{2}\log_2(2 \pi e(2 + \sigma^2)),
\end{equation}
from which it follows that
\begin{equation}
	\lim_{M \to \infty} \frac{I(\mathbf{y}; \mathbf{U},\mu)}{M} \leq \frac{1-\kappa}{2}\log\left(\frac{1}{\sigma^2}\right) + \frac{1}{2}\log_2(2+\sigma^2) 	- \frac{\kappa}{2}\log_2((\sqrt{1/\kappa}-1)^2 + \sigma^2).
\end{equation}
Applying the preceding to Lemma \ref{thm:fano}, we obtain the upper bound.

{\bf Lower Bound:}
Suppose that we choose $\Phi$ to be any matrix with orthonormal rows. We bound the pairwise misclassification error via Lemma \ref{thm:bhattacharyya}, which yields
\begin{multline}\label{eqn:affine.total.pairwise.error}
	\mathrm{P}_e(\mu_i,\mathbf{U}_i,\mu_j,\mathbf{U}_j) \leq
	\frac{1}{2} \cdot \left(\frac{|\frac{\Phi \mathbf{U}_i \mathbf{U}_i^T + \Phi \mathbf{U}_j \mathbf{U}_j^T\Phi^T + 2\sigma^2\mathbf{I}}{2}|}{|\Phi \mathbf{U}_i \mathbf{U}_i^T\Phi + \sigma^2 \mathbf{I}|^{\frac{1}{2}}|\Phi \mathbf{U}_j \mathbf{U}_j^T \Phi^T + \sigma^2 \mathbf{I}|^{\frac{1}{2}}}\right)^{-\frac{1}{2}} \cdot \\
	\exp \left(- \frac{1}{8} \cdot \left({\bf \mu}_i - {\bf \mu}_j\right)^T {\Phi}^T \left(\frac{ {\Phi} \left({\bf U}_i{\bf U}_i^T + {\bf U}_j{\bf U}_j^T\right) {\Phi}^T + 2 \sigma^2 {\bf I}}{2}\right)^{-1} {\Phi} \left({\bf \mu}_i - {\bf \mu}_j\right)\right).
\end{multline}
Observe that the argument of the exponential term is always nonnegative, so the exponential is always smaller than one. Therefore, the bound on the misclassification probability of affine subspaces is always smaller than that of linear subspaces, and the lower bound from Theorem \ref{thm:zero.mean.capacity} also applies to affine subspaces. Applying this fact yields the lower bound for $\kappa \geq 1/2$.

For $\kappa < 1/2$, we apply (\ref{eqn:zero.mean.pairwise.error}) to (\ref{eqn:affine.total.pairwise.error}), yielding
\begin{multline}
	\mathrm{P}_e(\mu_i,\mathbf{U}_i,\mu_j,\mathbf{U}_j) \leq 
	2^{\frac{k-2}{2}} \cdot \left(1 + \frac{(\sqrt{1/(2\kappa)}-1)^2 +\epsilon(M)}{\sigma^2}\right)^{-\frac{k}{2}} \cdot \\
	\exp \left(- \frac{1}{8} \cdot \left({\bf \mu}_i - {\bf \mu}_j\right)^T {\Phi}^T \left(\frac{ {\Phi} \left({\bf U}_i{\bf U}_j^T + {\bf U}_i{\bf U}_j^T\right) {\Phi}^T + 2 \sigma^2 {\bf I}}{2}\right)^{-1} {\Phi} \left({\bf \mu}_i - {\bf \mu}_j\right)\right),
\end{multline}
where again $\epsilon(M) \to 0$. Next, let $\Phi(\mathbf{U}_i\mathbf{U}_i^T + \mathbf{U}_j\mathbf{U}_j^T)\Phi^T = \mathbf{W}_{ij} \Lambda_{ij} \mathbf{W}_{ij}^T$ be the eigenvalue decomposition of the covariance pair sum.
Also define $\omega = \mathbf{W}_{ij}^T\Phi(\mu_i-\mu_j)/2$, which is i.i.d. Gaussian with zero mean and unit variance. We therefore obtain
\begin{equation*}
	\mathrm{P}_e(\mu_i,\mathbf{U}_i,\mu_j,\mathbf{U}_j) \leq 2^{\frac{k-2}{2}} \cdot \left(1 + \frac{(\sqrt{1/(2\kappa)}-1)^2 +\epsilon(M)}{\sigma^2}\right)^{-\frac{k}{2}} \cdot
	\exp \left(- \frac{1}{4}  \omega^T \left(\Lambda_{ij}/2 + \sigma^2 {\bf I}\right)^{-1} \omega\right).
\end{equation*}
With probability one, $\Lambda_{ij}$ contains $2k$ nonzero eigenvalues. The preceding bound increases in these eigenvalues, so to bound the error we bound the eigenvalues by infinity, which yields
\begin{equation*}
	\mathrm{P}_e(\mu_i,\mathbf{U}_i,\mu_j,\mathbf{U}_j) \leq 2^{\frac{k-2}{2}} \cdot \left(1 + \frac{(\sqrt{1/(2\kappa)}-1)^2 +\epsilon(M)}{\sigma^2}\right)^{-\frac{k}{2}} \cdot
	\prod_{i=2k+1}^{M} \exp\left(-\frac{1}{4\sigma^2}\omega_i^2\right).
\end{equation*}
Taking the expectation yields
\begin{equation*}
	E[\mathrm{P}_e(\mu_i,\mathbf{U}_i,\mu_j,\mathbf{U}_j)]	\leq  2^{\frac{k-2}{2}} \cdot \left(1 + \frac{(\sqrt{1/(2\kappa)}-1)^2 +\epsilon(M)}{\sigma^2}\right)^{-\frac{k}{2}} \cdot
	\prod_{i=2k+1}^{M} E\left[\exp\left(-\frac{1}{4\sigma^2}\omega_i^2)\right)\right],
\end{equation*}
where the expectation moves inside the product because each $\omega_i$ is independent of the others. Noting that each expectation in the final expression is just the moment-generating function of a Chi-squared random variable, we obtain
\begin{equation}
	E[\mathrm{P}_e(\mu_i,\mathbf{U}_i,\mu_j,\mathbf{U}_j)]	\leq  2^{\frac{k-2}{2}} \cdot \left(1 + \frac{(\sqrt{1/(2\kappa)}-1)^2 +\epsilon(M)}{\sigma^2}\right)^{-\frac{k}{2}} \cdot
	\left(1 +\frac{1}{2\sigma^2}\right)^{-\frac{M-2k}{2}} \label{eqn:affine.pairwise.probability}
\end{equation}
Applying the union bound and taking the logarithm, we obtain
\begin{align}
	\log_2(E[\mathrm{P}_e(\mathbf{a})]) &\leq \rho M - \frac{k}{2}\log_2\left(1 + \frac{(\sqrt{1/(2\kappa)}-1)^2 +\epsilon(M)}{\sigma^2} \right) - \frac{M-2k}{2}\log_2\left( 1+\frac{1}{2 \sigma^2} \right) + \frac{k-2}{2} \\
	&\leq \rho M - \frac{M-k}{2}\log_2\left(1 + \frac{\min\{(\sqrt{1/(2\kappa)}-1)^2 +\epsilon(M),1/2\}}{\sigma^2} \right) + \frac{k-2}{2}.
\end{align}
Letting $M \to \infty$, we obtain the lower bound for $\kappa < 1/2$.
\end{IEEEproof}

For affine subspaces, the bounds are tight to within an $O(1)$ term for all values of $\kappa$. Roughly speaking, the term in the Bhattacharyya bound associated with discriminating the means cancels out the gap to the upper bound associated with discriminating the associated linear subspaces. Therefore pairwise analysis, along with the union bound, is sufficient for establishing tight bounds on the classification capacity for affine subspaces even when it fails for linear subspaces.


\section{Diversity-discrimination Tradeoff}\label{sect:ddt}
Here we prove bounds on the diversity-discrimination tradeoff. Similar to the classification capacity, we prove an upper bound on $d_\mathrm{linear}(r)$, which shows that the maximum diversity gain is $\min\{k,M-k\}$ and the maximum discrimination gain is $M-k$. We also prove a lower bound, based on the Bhattacharyya bound, which establishes that the average misclassification probability decays at least as $(1/\sigma^2)^{-\frac{\min\{k,M-k \}-r}{2}}$ when the number of subspaces grows as $(1/\sigma^2)^{\frac{r}{2}}$. For affine subspaces, we prove an upper bound which shows that the misclassification probability decays no faster than $(1/\sigma^2)^{-\frac{M-k-r}{2}}$. In this case, the Bhattacharyya analysis shows that the upper bound is tight.
\subsection{Linear Subspaces}\label{sect:zero.mean.ddt}
First, we prove bounds on $d_\mathrm{linear}(r)$.
\begin{theorem}\label{thm:zero.mean.ddt}
	For linear subspaces, the DDT is upper bounded by
	\begin{equation}\label{eqn:linear.ddt.upper.bound}
		d_\mathrm{linear}(r) \leq \left[\min\left\{M-k-r, k\left(1-\frac{r}{M}\right)\right\}\right]^+,
	\end{equation}
	and the DDT is bounded below by
	\begin{equation}\label{eqn:zero.mean.ddt.lower.bound}
		d_\mathrm{linear}(r) \geq \left[\min\{M-k,k\} - r\right]^+.
	\end{equation}
\end{theorem}
\begin{IEEEproof}
First we prove the upper bound, the first term of which follows from Lemma \ref{thm:fano.ddt}. Combining (\ref{eqn:zero.mean.conditional.entropy}) and (\ref{eqn:zero.mean.marginal.entropy}), it is easy to see that
\begin{equation}
	\lim_{\sigma^2 \to 0} \frac{I(\mathbf{U}; \mathbf{y})}{\frac{1}{2}\log(1/\sigma^2)} \leq M-k.
\end{equation}
Therefore, by Lemma \ref{thm:fano.ddt}, $d_\mathrm{linear}(r)=0$ whenever $r \geq M-k$.
Next, suppose that $d_\mathrm{linear}(r) > M-k-r$ for some $0 \leq r < M-k$, meaning that, for some $\epsilon > 0$,
\begin{equation}\label{eqn:arbitrary.ddt.pe}
	\log_2(\bar{\mathrm{P}}_e(r,\sigma^2)) \leq -\frac{M-k-r+\epsilon}{2}\log_2(1/\sigma^2) + o(\log(\sigma^2)).
\end{equation}
Using the union bound, we can express the probability of error for $r=M-k$ in terms of (\ref{eqn:arbitrary.ddt.pe}):
\begin{align*}
	\log_2(\bar{\mathrm{P}}_e(M-k,\sigma^2)) &\leq  \log_2((1/\sigma^2)^{\frac{M-k-r}{2}} \bar{\mathrm{P}}_e(r,\sigma^2)) \\
	&\leq \frac{M-k-r}{2} \log_2(1/\sigma^2) + \log_2(\bar{\mathrm{P}}_e(r,\sigma^2)) \\
	&\leq -\frac{\epsilon}{2}\log_2(1/\sigma^2) + o(\log(\sigma^2)).
\end{align*}
This implies $d_\mathrm{linear}(M-k) > 0$, which is a contradiction.

The second term in the upper bound follows from an ``outage''-style argument reminiscent of that of \cite{zheng:Allerton02}. For linear subspaces, we can rewrite the signal model (\ref{eqn:signal.model}) as
\begin{equation*}
	\mathbf{y} = \Phi \mathbf{U} \mathbf{h} + \mathbf{z},
\end{equation*}
where $\mathbf{h} = (h_1,\cdots h_k)^T \sim \mathcal{N}(0,\mathbf{I})$. We define an outage event
\begin{equation}
	\mathcal{F} = \{ h_i^2 \leq (1/\sigma^2)^{-\beta}, \forall \ i\},
\end{equation}
for $0 \leq \beta \leq 1$. Because each $h_i^2$ is Chi squared with a single degree of freedom,
\begin{equation}
	\mathrm{Pr}(\mathcal{F}) \leq (1/\sigma^2)^{-\frac{k\beta}{2}} \cdot \exp(k/2).
\end{equation}
Next, we bound the {\em conditional} normalized mutual information:
\begin{align*}
	\lim_{\sigma^2 \to 0} \frac{I(\mathbf{U} ; \mathbf{y} | \mathcal{F})}{1/2\log_2(1/\sigma^2)} &= \lim_{\sigma^2 \to 0} \frac{h(\mathbf{y} | \mathcal{F}) - h(\mathbf{y} | \mathbf{U},\mathcal{F}) }{1/2\log_2(1/\sigma^2)} \\
	&\leq \lim_{\sigma^2 \to 0} \frac{\log_2(E_{\mathbf{h},\mathbf{U}}[\det(\Phi\mathbf{U}\mathbf{h}\mathbf{h}^T\mathbf{U}^T\Phi^T + \sigma^2\mathbf{I}) | \mathcal{F}]) - M/2\log_2(\sigma^2) }{1/2\log_2(1/\sigma^2)},
\end{align*}
where the inequality follows because (i) the Gaussian distribution maximizes mutual information, and (ii) $h(\mathbf{y} | \mathbf{U},\mathcal{F})) \geq h(\mathbf{z})$ by the entropy power inequality. Conditioned on the outage event, we have $\mathbf{h}\mathbf{h}^T \leq (1/\sigma^2)^{-\beta} \cdot \mathbf{I}$, from which it follows that
\begin{align*}
	\lim_{\sigma^2 \to 0} \frac{I(\mathbf{U} ; \mathbf{y} | \mathcal{F})}{1/2\log_2(1/\sigma^2)} &\leq \lim_{\sigma^2 \to 0} \frac{\log_2(E_{\mathbf{U}}[|(1/\sigma^2)^{-\beta} \cdot \Phi\mathbf{U}\mathbf{U}^T\Phi^T + \sigma^2\mathbf{I}|]) - M/2\log_2(\sigma^2) }{1/2\log_2(1/\sigma^2)}  \\
	&\leq \lim_{\sigma^2 \to 0} \frac{M/2\log_2((1/\sigma^2)^{-\beta} + \sigma^2) - M/2\log_2(\sigma^2) }{1/2\log_2(1/\sigma^2)} \\
	&= M(1-\beta).
\end{align*}
By the law of total probability,
\begin{align*}
	E[\mathrm{P}_e(\mathbf{a})] &= E[\mathrm{P}_e(\mathbf{a}) | \mathcal{F}] \mathrm{Pr}(\mathcal{F}) + E[\mathrm{P}_e(\mathbf{a}) | \mathcal{F}^c](1- \mathrm{Pr}(\mathcal{F})) \notag \\
	&\geq E[\mathrm{P}_e(\mathbf{a}) | \mathcal{F}] \mathrm{Pr}(\mathcal{F}). \label{eqn:outage.mi}
\end{align*}
By Lemma \ref{thm:fano.ddt}, whenever $r > M(1-\beta)$, the conditional probability $E[P(\mathbf{a}) | \mathcal{F}]$ is bounded away from zero. Therefore,
\begin{align*}
	d_\mathrm{linear}(M(1-\beta) + \epsilon) &= \lim_{\sigma^2 \to 0} -\frac{\log_2(E[\mathrm{P}_e(\mathbf{a})])}{1/2\log_2(1/\sigma^2)} \\
		&\leq \lim_{\sigma^2 \to 0} -\frac{ \log_2(E[P(\mathbf{a}) | \mathcal{F}]) + \log_2(\mathrm{Pr}(\mathcal{F}))}{1/2\log_2(1/\sigma^2)} \\
	&= \lim_{\sigma^2 \to 0} -\frac{\log_2(\mathrm{Pr}(\mathcal{F}))}{1/2\log_2(1/\sigma^2)} \\
	&\leq k \beta.
\end{align*}
Taking $\epsilon \to 0$, we obtain the second term of the upper bound.

The lower bound follows from the Bhattachayya analysis from Section \ref{sect:zero.mean.capacity}. For discrimination gain $r$, we combine (\ref{eqn:zero.mean.pairwise.error}) with the union bound, yielding
\begin{equation*}
	\log_2(\bar{\mathrm{P}}_e(r,\sigma^2)) \leq \frac{r}{2}\log_2\left(\frac{1}{\sigma^2}\right) - \frac{\min\{M-k,k\}}{2}\log_2\left(1 + \frac{(\sqrt{1/\kappa} - \sqrt{2})^2 +\epsilon(M)}{\sigma^2}\right) + \frac{k}{2}.
\end{equation*}
Therefore, we have
\begin{equation*}
	d_\mathrm{linear}(r)= \lim_{\sigma^2 \to 0} \frac{\log_2(\bar{\mathrm{P}}_e(r,\sigma^2))}{\frac{1}{2}\log_2(1/\sigma^2)} \geq \min\{M-k,k\} - r.
\end{equation*}
\end{IEEEproof}

Similar to the classification capacity, the lower bound is tight when $k \geq M/2$. Otherwise, the lower bound achieves full diversity for $r=0$, but falls short of the upper bound for higher discrimination gain. Note, however, that the second term in (\ref{eqn:linear.ddt.upper.bound}), which establishes that the diversity gain is no greater than $k$, is clearly loose because it predicts nonzero diversity for discrimination gains higher than $M-k$. This looseness is due to the bound on the normalized mutual information, in which we employed $h(\mathbf{y} | \mathbf{U},\mathcal{F})) \geq h(\mathbf{z})$; this bound neglects the effect of the outage event on the eigenvalues.

A tighter bound on the conditional entropy is difficult because $\mathbf{y}$ is no longer Gaussian conditioned on $\mathcal{F}$. However, we can make heuristic calculations by bounding the conditional covariance and supposing that the entropy is approximately that of the equivalent Gaussian. Then, the normalized mutual information is instead bounded by $(M-k)(1-\beta)$. Following that analysis leads to the following bound on DDT function
\begin{equation}\label{eqn:linear.ddt.conjecture}
	d_\mathrm{linear}(r) = \min\{M-k,k\}\left[1 - \frac{r}{M-k} \right]^+.
\end{equation}
This function is just the line segment connecting the maximum diversity order and the maximum discrimination gain in the upper bound. Based on the preceding intuition and the numerical results in Section \ref{sect:experiments}, we conjecture that this is the true diversity-discrimination tradeoff for linear classification.

\subsection{Affine Subspaces}
Next, we derive $d_\mathrm{affine}(r)$.
\begin{theorem}\label{thm:nonzero.mean.ddt}
For affine subspace classification, the DDT is
\begin{equation}
	d_\mathrm{affine}(r) = [M - k - r]^+.
\end{equation}
\end{theorem}
\begin{IEEEproof}
We can upper bound the DDT using the same argument as in the proof of Theorem \ref{thm:zero.mean.ddt}. Combining (\ref{eqn:affine.conditional.entropy}) and (\ref{eqn:affine.marginal.entropy}), we obtain
\begin{equation}
	\lim_{\sigma^2 \to 0} \frac{I(\mu,\mathbf{U}; \mathbf{y})}{\frac{1}{2}\log(1/\sigma^2)} \leq M-k.
\end{equation}
Therefore, $d_\mathrm{affine}(r) = 0$ for $r \geq M-k$ by Lemma \ref{thm:fano.ddt}. As before, by the union bound there is a contradiction if $d(r) > M-k-r$ for any $0 \leq r \leq M-k$. 

To lower bound the DDT, observe from the proof of Theorem \ref{thm:nonzero.mean.capacity} that
\begin{equation}
	\log_2(E[\mathrm{P}_e(\mu_i,\Sigma_i,\mu_j,\Sigma_j)]) \leq -\frac{M-k}{2}\log_2(1/\sigma^2) + o(\log(\sigma^2)).
\end{equation}
For discrimination gain $0 \leq r \leq M-k$, the union bound yields
\begin{equation}
	\log_2(\bar{\mathrm{P}}_e(r,\sigma^2)) \leq -\frac{M-k-r}{2}\log_2(1/\sigma^2) + o(\log(\sigma^2)),
\end{equation}
which establishes the result.
\end{IEEEproof}

For affine spaces, similar to the classification capacity, the upper bound is tight. Therefore, the Bhattacharyya bound is tight not only with respect to the pairwise error, but also upon application of the union bound. Therefore, while the translation of linear subspaces into affine subspaces does not necessarily improve the {\em number} of subspaces that a classifier can discriminate reliably, for $k < M/2$ translation does improve the decay of the error probability as the noise power vanishes.

\section{Numerical Results}\label{sect:experiments}
Here we validate our results numerically. First, we study the performance of linear subspace classifiers with subspaces drawn randomly from $p_\mathrm{linear}$, focusing on the regimes in which the upper and lower bounds disagree and drawing conclusions about the tightness of the bounds. Then, we study the classifier performance over the YaleB database of face images, comparing empirical performance to the predictions of Section \ref{sect:capacity}.

\subsection{Linear Subspaces}
Here, we want to see whether the upper bound on the $C_\mathrm{linear}(\nu,\kappa)$ is tight for $\kappa < 1/2$, as we conjectured in Section \ref{sect:zero.mean.capacity}. We also want to see whether the diversity-discrimination function we conjectured in Section \ref{sect:zero.mean.ddt} is correct.

To answer these questions, we examine classifier performance as $\sigma^2 \to 0$. We draw subspaces from $p_\mathrm{linear}$ and we choose $\Phi$ to be the first $M$ rows of a randomly-chosen unitary matrix. We corrupt the features with white Gaussian noise of variance $\sigma^2$, and we perform maximum-likelihood classification on the noisy features. Because of computational limitations, it is infeasible to study empirical performance as the signal dimension becomes large. Therefore, instead of testing the classification capacity directly, we examine the DDT performance. If, for discrimination gain $r$, the diversity gain is nonzero, then the classification capacity must be at least as great as $r/M \log_2(1/\sigma^2) + o(\log(\sigma^2))$.

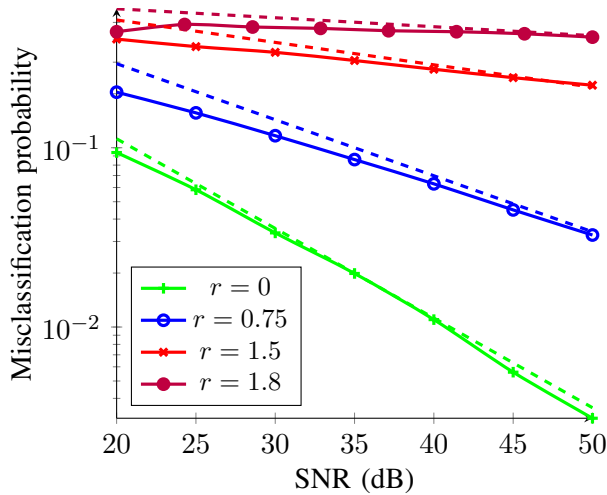
\begin{figure}[htbp]
\begin{center}
	\begin{tikzpicture}
	
	\begin{semilogyaxis}[
	xlabel={SNR (dB)},
	ylabel={Misclassification probability},
	axis x line=bottom,
	axis y line=left,
	height=200pt,
	width=225pt,
	legend pos = south west,
	]

		\addplot[smooth,color=green,very thick,solid,mark=+] file {plots/linear310.dat};
		\addlegendentry{\small $r=0$};
		
		\addplot[smooth,color=blue,very thick,solid,mark=o] file {plots/linear310.75.dat};
		\addlegendentry{\small $r=0.75$};
		
		\addplot[smooth,color=red,very thick,solid,mark=x] file {plots/linear311.5.dat};
		\addlegendentry{\small $r=1.5$};
		
		\addplot[smooth,color=purple,very thick,solid,mark=*] file {plots/linear311.8.dat};
		\addlegendentry{\small $r=1.8$};
		
		\addplot[smooth,color=green,very thick,dashed] file {plots/linear310.theory};
		
		\addplot[smooth,color=blue,very thick,dashed] file {plots/linear310.75.theory};
		
		\addplot[smooth,color=red,very thick,dashed] file {plots/linear311.5.theory};
		
		\addplot[smooth,color=purple,very thick,dashed] file {plots/linear311.8.theory};
		
	\end{semilogyaxis}
\end{tikzpicture}
\caption{Misclassification probability vs. SNR for linear subspaces. Dashed lines indicate the slopes predicted by (\ref{eqn:linear.ddt.conjecture}).}
\label{fig:linear.simulations}
\end{center}
\end{figure}

In Figure \ref{fig:linear.simulations} we plot the misclassification probability as a function of $\sigma^2$. For each value of $\sigma^2$, we compute the average misclassification probability over $10^2$ realizations of the subspaces $\mathbf{U}_l$ and $10^2$ realizations of the signal of interest per set of subspaces. We also plot the error slopes predicted by (\ref{eqn:linear.ddt.conjecture}). We select dimensions $N=M=3$, $k=1$ and discrimination gains $r \in \{0, 0.75, 1.5, 1.8\}$. We observe decaying misclassification probability for all values of $r$; furthermore, we observe rates of decay consistent with the conjectured DDT function. Therefore, we conclude that, regardless of $\kappa$, the classification capacity satisfies
\begin{equation*}
	C_\mathrm{linear}(\nu,\kappa) = \frac{1-\kappa}{2}\log_2(1/\sigma^2) + o(\log_2(\sigma^2)),
\end{equation*}
and that the DDT is
\begin{equation*}
	d_\mathrm{linear}(r) = \min \{k,M-k \}\left[ 1 - \frac{r}{M-k} \right]^+.
\end{equation*}
Recall that these conclusions were proven only in the regimes $k \geq M/2$ or $\kappa \geq 1/2$.

\subsection{Face Recognition}
Next, we explore the correspondence between the theoretical results derived in the previous sections and a practical face recognition application. We examine face recognition when the orientation of the face relative to the camera remains fixed but the illumination varies. Supposing the faces to be approximately convex and to reflect light according to Lambert's law, \cite{basri:PAMI03} shows via spherical harmonics that the set of images of an individual face lies approximately on a nine-dimensional subspace, regardless of the inherent dimension of the images. It is therefore sufficient to discriminate between the subspaces to classify faces.

\begin{figure}[htbp]
\begin{center}
	\includegraphics[width=100px]{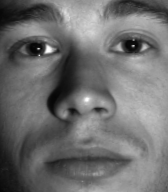}
	\includegraphics[width=100px]{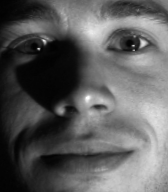}
\caption{Two sample images from the Extended Yale Face Database B. These images are of the same face, but are taken under different illumination conditions.}
\label{fig:faces}
\end{center}
\end{figure}

We use 38 cropped faces from the Extended Yale Face Database B, described in \cite{georghiades:PAMI01,lee:PAMI05}. For each face, the database contains a few dozen greyscale photographs, each having 32,256 pixels, taken under a variety of illumination conditions as shown in Figure \ref{fig:faces}. We vectorize these images and pass them through a feature matrix $\Phi$, chosen as before to be the first $M$ rows of an arbitrary unitary matrix. We classify the faces using the maximum-likelihood classifier supposing zero-mean Gaussian classes. We divide the database into two, using half of the images to estimate the nearest covariance for each face, using the other half as test images.

\begin{figure}[htbp]
\begin{center}
	\includegraphics[width=250px]{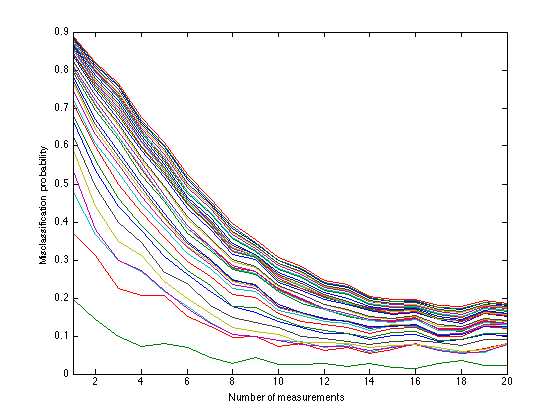}
\caption{Misclassification probability as a function of $M$, for $L$ ranging from 1 to 38.}
\label{fig:error.performance}
\end{center}
\end{figure}

In Figure \ref{fig:error.performance} we plot the misclassification probability as a function of $M$ and for $L$ ranging from 2 to 38. While we do not label each curve, it is easy to see that the misclassification probability increases with $L$ and decreases with $M$. However, even for large $M$ the error probability remains as high as 0.2 for $L=38$. We take 0.2 as a baseline for ``successful'' performance when the number of faces and signal dimension are high.

Finally, we examine how well our theory predicts the performance seen here. To estimate the noise power $\sigma^2$, we project each image onto its estimated subspace, transformed by $\Phi$, and we take take the projected squared norm as the signal power and the squared residual norm, normalized by the number of features $M$, as the noise power. We then estimate the number of faces that Theorem \ref{thm:zero.mean.capacity} predicts can be discriminated reliably. Discarding the constants, we simply compute
\begin{equation}
	\max\{1,\min\{\mathrm{1/\sigma^2}^{(M-9)/2},38\}\}.
\end{equation}
Naturally, this number grows quickly in $M$, and beyond $M=11$ or $M=12$, theory suggests that we ought to be able to discriminate all 38 of the faces with low probability of error. In Figure \ref{fig:number.of.classes} we compare this prediction against the empirical performance of our classifier. Using the results shown in Figure \ref{fig:error.performance}, we compute, for each $M$, the maximum $L$ for which the probability of error is less than 0.2.

\begin{figure}[htbp]
\begin{center}
	\begin{tikzpicture}
	
	\begin{axis}[
	xlabel={M},
	ylabel={Number of discernible faces},
	axis x line=bottom,
	axis y line=left,
	height=200pt,
	width=225pt,
	legend pos = north west,
	]

		\addplot[smooth,color=green,very thick,solid,mark=*] file {plots/face.dat};
		\addlegendentry{\small Observed};
		
		\addplot[smooth,color=blue,very thick,dashed] file {plots/face.theory};
		\addlegendentry{\small Predicted};
		
	\end{axis}
\end{tikzpicture}	
\caption{Misclassification probability as a function of $M$, for $L$ ranging from 1 to 38.}
\label{fig:number.of.classes}
\end{center}
\end{figure}

The empirical performance is similar to theoretical prediction. As $M$ increases past 9, the number of faces rises swiftly as predicted. After $M=15$ or so, all 38 of the faces can be discriminated, and it is not advantageous to extract more features. We do observe, however, that the transition is not as sharp as Theorem \ref{thm:zero.mean.capacity} predicts. Whereas the theoretical transition occurs over only 2-3 features, in practice the transition stretches out over 5-10 features. In addition to mild model mismatch due to non-Lambertian reflectances, shadows due to the non-convexity of real faces, imperfect estimation of subspaces, etc., we suspect that this is primarily a phenomenon of classification at finite dimension. The transition between success and failure becomes sharp in the limit, but remains gradual when dimensions measure in the tens or hundreds.

\section{Conclusion}
Inspired by dualities between wireless communication over non-coherent channels and the classification from noisy, linear features, we have derived fundamental limits on the classification of linear and affine subspaces from noisy, linear features. We defined performance limits reminiscent of those in wireless information theory: the classification capacity, which governs classifier performance in the limit of high signal dimension, and the diversity-discrimination tradeoff, which governs classifier performance in the limit of low noise power. We proved inner and outer bounds on these quantities. For linear subspaces, the bounds are tight in some regimes of $N,M$, and $k$, and for affine subspaces they are tight everywhere. Based on numerical evaluation, we conjectured that the true classification capacity and DDT for linear subspaces in the regimes in which the bounds are not tight. Beyond the characterization of such limits, we showed via an application to face recognition that theoretical trends agree reasonably with practical ones.

\bibliography{/Users/nokleby/documents/LaTeX/bibliography}

\end{document}